\documentstyle[aps,multicol]{revtex}
\newcommand{\beq}{\begin{equation}} \newcommand{\eeq}{\end{equation}}
\newcommand{\bea}{\begin{eqnarray}} \newcommand{\eea}{\end{eqnarray}}
\newcommand{\bear}{\begin{eqnarray*}} \newcommand{\eear}{\end{eqnarray*}}
 \newcommand{\rf}[1]{(\ref{#1})}

\begin{document}

\title {Frontier between separability and quantum entanglement in a many spin system}

\author{Francisco C. Alcaraz $^{(a)}$ and Constantino Tsallis $^{(b,c)}$}

\address{$^{(a)}$ Departamento de F\'{\i}sica, Universidade Federal de S\~ao Carlos, Via Washington Luiz, S\~ao Carlos-SP, Brazil\\
$^{(b)}$ Centro Brasileiro de Pesquisas F\'{\i}sicas, Xavier Sigaud 150, 22290-180, Rio de Janeiro-RJ, Brazil (tsallis@cbpf.br)\\
$^{(c)}$ Erwin Schroedinger International Institute for Mathematical Physics, Boltzmanngasse 9, A-1090 Wien, Austria}

\date{\today}

\maketitle

\begin{abstract}

We discuss the critical point $x_c$ separating the 
quantum entangled and separable states in two series of 
 $N$ spins $S$ in the simple mixed state 
characterized by the matrix operator 
$\rho=x|\tilde{\phi}$$><$$\tilde{\phi}| + \frac{1-x}{D^N} I_{D^N}$ 
where $x \in [0,1]$, $D =2S+1$, ${\bf I}_{D^N}$ is the $D^N \times 
D^N$ unity matrix  and $|\tilde {\phi}> $ 
is a  special entangled state.  
The cases $x=0$ and $x=1$ correspond respectively to fully random spins and to a fully entangled state.
In the first of these series we consider special states $|\tilde{\phi}>$ invariant under 
charge conjugation, that generalizes the $N=2$ spin $S=\frac{1}{2}$ Einstein-Podolsky-Rosen state, 
and in the second one we consider generalizations of the Weber density matrices. 
The evaluation of the critical point $x_c$ was done through bounds coming 
from the partial transposition method of Peres and the conditional 
nonextensive entropy criterion. Our results suggest the conjecture that whenever 
the bounds coming from both methods coincide the result of $x_c$ is the exact one. 
The results we present are relevant for the discussion of quantum computing, 
teleportation 
and cryptography.

\end{abstract}

\begin{multicols}{2} 

Quantum entanglement is a quite amazing physical phenomenon. It was first discussed in depth in 1935 by Einstein, Podolsky and Rosen (EPR) \cite{EPR} and by Schroedinger \cite{schroedinger}, and since then by many others\cite{GHZ,ekert,bennett,zurek2,quantumchaos,popescu,barenco,zurek,horodecki1,horodecki2,peres,horod,horodeckicube,popescu2,horodecki3,lloyd,bruss,zeilinger}. This phenomenon has attracted intensive interest in recent years due to its applications in quantum computation, teleportation and cryptography, as well as to its connections to quantum chaos. Special importance is attributed to the frontier which, in mixed states, separates the quantum entangled from the so called {\it separable} states. Separability, precisely defined in what follows, is essentially a classical concept, and it is the one on which the possibility of a description in terms of local realism with hidden variables is currently built on. The state of a system composed of subsystems $A$ and $B$ is said separable (or unentangled) if and only if its corresponding density matrix $\rho_{A+B}$ can be written as
\begin{equation}
\rho_{A+B} = \sum_{i=1}^W p_i \;\rho_A^{(i)} \otimes \rho_B^{(i)} \;\;\;\; (p_i \ge 0\; \forall i; \;\sum_{i=1}^W p_i=1).
\end{equation}
The frontier between quantum entangled and separable states for a variety of mixed 
states is currently discussed in the literature. 
Several {\it necessary} conditions for separability are already available; however, no general {\it sufficient} conditions are yet known. 
Among the existing necessary conditions two criteria have shown some 
preference in the literature due to their simplicity and the fact that in 
some cases these conditions coincide with the exact ones. 
Those are the Peres criterium\cite{peres} based on the positivity of the 
partially transposed density matrix, and the generalization of  
standard entropic methods, given by the conditional nonextensive 
entropy criterium, recently advanced by Abe and 
Rajagopal\cite{aberajagopal}.
In the present paper we are going to apply and compare both methods by applying them to 
several distinct density matrices coming from some many body systems.
Both criteria recover exact separable-entangled frontiers 
in some simple mixed states of a two spin $1/2$ 
system \cite{aberajagopal,sethmichel}, 
but only provide a bound in more 
complex cases.
Although no rigorous proof is available, as we shall see, the analysis 
of some particular cases suggests that generically the partial transpose 
method overestimates the separable region either {\it equally} or {\it less} 
than the conditional entropy method. Moreover, our results indicate 
that whenever both methods yield the same result, this result is the exact one. 
It is clear that the discussion of quantum entanglement in many body 
systems is crucial both conceptually (e.g., for understanding the 
interface between the classical and quantum descriptions of the 
real world) and practically (e.g., for efficiently implementing 
the input and output of a device such as a quantum computer). 
Some specific $N$-body systems have already been discussed in the 
literature (see \cite{michalpawel,abe2001} for some classes 
of systems with arbitrary $N$, and \cite{rajagopalrendell} 
for the Greenberger-Horne-Zeilinger state of $N=3$ spins $1/2$; 
see \cite{terhal} for a review).

Let us consider a $N$-body system composed by spin-$S$ operators 
acting on a Hilbert space $H=H_1 \otimes \ldots \otimes H_n$ of 
dimension $D^N = (2S+1)^N$. An arbitrary density matrix acting 
on this space should have all its eigenvalues non negative. 
Consider  an 
arbitrary orthogonal basis $|s_1,\ldots,s_N>$ ($s_i =-S,\ldots,S$) 
spanning the Hilbert space. The Peres criterion appropriate to 
this $N$-body systems asserts that from a given density matrix 
$\rho$, with elements $<s_1,\ldots,s_N|\rho |s_1',\ldots,s_N'>$, 
it is possible to define other density matrices 
$\sigma^{i_1,\ldots,i_n}$, where the spin coordinates $i_1,\ldots,i_n $ 
($i_1,\ldots,i_n =1,\ldots,N$), are transposed ("in $\leftrightarrow $ 
out "), i. e., 
$<s_1 \cdots s_N| \sigma^{i_1,\ldots,i_n} 
|s_1' \cdots s_N'> =$ 
\bea \label{AA5}
&& <s_1 \cdots s_{i_1-1}s_{i_1}' s_{i_1+1}
 \cdots s_{i_2}' \cdots 
s_{i_n}' s_{i_n+1} \cdots s_N| \nonumber \\ && \rho 
 |s_1' \cdots s_{i_1-1}'s_{i_1} s_{i_1+1}' \cdots s_{i_2} \cdots 
s_{i_n} s_{i_n+1}' \cdots s_N'>. \nonumber  
\eea
The non-negativity of these new density matrices give us the 
Peres criterion, namely all eigenvalues of $\sigma^{i_1,\ldots,i_n}$ should be
non-negative.

In order to probe and compare the above mentioned methods  we 
will consider several sets of density matrices of general type 

\beq \label{A2} 
\rho(x) = (1-x)\rho(0) + x \tilde{\rho}
\eeq
where 
$\rho(0) = \rho_1\otimes \ldots \otimes \rho_N = 
{\bf I}_{D^N}/D^N$, ${\bf I}_{D^N}$ being the $D^N \times 
D^N$ unity matrix,  
 is a fully separable state 
and $\tilde \rho = |\tilde \phi > <\tilde \phi|$ is a density 
matrix corresponding to a known entangled state $|\tilde \phi>$. 
Then as we increase the value of $x$, there exists a critical value $x = x_c$
where the  quantum entanglement takes place, and we want to obtain 
bounds for $x_c$. 

The matrices 
$\sigma^{i_1,\ldots ,i_n}$ obtained by the partial transposition 
at indexes ($i_1,\ldots,i_n$), are given by 

\beq \label{A3}
{\sigma}^{i_1,\cdots,i_n}(x) = \frac{1-x}{D^N} {\bf I}_{D^N} + 
x\tilde{\sigma}^{i_1,\ldots,i_n}
\eeq
where $<s_1,\ldots,s_N|\tilde \sigma^{i_1,\ldots,i_n}|s_1',\ldots,
s_N'> = <s_1,\ldots,s_{i_1}' \cdots s_{i_2}' \cdots,
s_{i_n}',\ldots,s_N| \tilde \rho 
|s_1' \cdots s_{i_1} \cdots s_{i_2} \cdots s_{i_n} \\ 
\cdots s_{N}'>$. 
The $D^N$ eigenvalues of 
$\sigma^{i_1,\ldots,i_n} (x)$ are given by
$E_i = (1-x)/D^N + x\tilde E_i$, where $\tilde E_i$ ($i=1,\ldots,D^L$) are 
the eigenvalues of $\tilde \sigma^{i_1,\ldots,i_n}$. The condition 
 for non-negative eigenvalues imply that quantum entanglement is present at least for $x \ge x_c^P$ ($P$ stands for Peres), where
\beq \label{AA4}
x_c^P = \frac{1}{1 -D^N{\tilde{\tilde E}}},
\eeq
and $\tilde{\tilde E}$ is the lowest eigenvalue of 
$\tilde \sigma^{i_1,\ldots i_n}$. 
In principle we should consider all the possible transpositions 
${i_1,\ldots,i_n}$ but, in all the applications we considered, the 
transposition of a single spin, i. e., $\sigma^{i_1}$, gave us the 
best (i.e., the lowest) bound for $x_c$. 

The second criterion we analyze was introduced recently by Abe and
Rajagopal\cite{aberajagopal}. Exploring the classical composition of 
probabilities arising from the nonextensive 
entropic form on which nonextensive statistical mechanics  \cite{tsallis} is based, they provide a condition for quantum unentanglement. 
Within this entropic approach the condition for quantum unentanglement 
is that the conditional probability for any subsystem forming 
the whole system should be non-negative. In this entropic approach 
the entropy of a $N$-spin $S$ system $A_1+ \cdots A_N$, formed by 
subsystems $A_i$ ($i=1,\ldots,N$), with density $\rho_{A_1+\cdots A_N}$ is 
given by
\beq \label{A4}
S_q(A_1+\cdots A_N) = \frac{1 - \mbox{Tr}\rho_{A_1+\cdots+A_N}^q}{q-1},
\eeq
where $q$ is an arbitrary real number. 
The particular case $q \rightarrow 
1$ recovers the standard von Neumann entropy. 
The conditional entropies, 
taking into account a partial information of the 
system, are given by
\bea \label{A5}
&&S_q(A_1+\cdots +A_{i_1-1}+A_{i_1+1}+\cdots+A_{i_n-1}+A_{i_n+1}+
\nonumber \\
&&\cdots +A_N| 
A_{i_1}+A_{i_2}+\cdots +A_{i_n}) = \nonumber \\
&&\frac{S_q(A_1+\cdots+A_N) -
S_q(A_{i_1}+A_{i_2}+\cdots +A_{i_n})}
{1 +(1-q)S_q(A_{i_1}+A_{i_2}+\cdots +A_{i_n})}
\eea
where 
$S_q(A_1+\cdots + A_n)$ is given by \rf{A4} with  
\bea \label{A6}
&&\rho_{A_{i_1}+\cdots +A_{i_n}} = \nonumber \\
&& \mbox{Tr}_{A_1,\ldots,A_{i_1-1},A_{i_1+1},\ldots,
A_{i_{n-1}},A_{i_{n+1}},\ldots,A_N} 
[\rho_{A_1+\cdots +A_N}].
\eea
The condition for non-negative conditional probability imply that, for 
arbitrary combinations $\{i_1,\ldots,i_n\}$, \rf{A5} should be non-negative, 
and consequently the bound for $x_c$ is evaluated by 
imposing 
\beq \label{A7}
\mbox{Tr}\rho_{A_1+\cdots A_N}^q = \mbox{Tr}\rho_{A_{i_1}+\cdots +A_{i_n}}^q.
\eeq
The left side of the above equation is simple to calculate for the special 
set of density matrices \rf{A2}
  by going to the basis where $\tilde \rho$ is 
diagonal, and we obtain 
\beq \label{A8}
\mbox{Tr}\rho_{A_1 +\cdots +A_n}^q = 
(D^N-1)(\frac{1-x}{D^N})^q + (\frac{1+(D^N-1)x}{D^N})^q \nonumber.
\eeq
From \rf{A2} and \rf{A6} we obtain
\beq \label{A9}
\rho_{A_{i_1}+\cdots + A_{i_n}} = \frac{1-x}{D^N} D^{N-n} 
{\bf I}_{D^n} + x \tilde{\tilde \rho}_{A_{i_1}+\cdots+A_{i_n}},
\eeq
where
\bea \label{C6}
\tilde{\tilde \rho}_{A_{i_1}+\cdots +A_{i_n}} =&& 
\mbox{Tr}_{A_{i_1},\ldots,A_{i_1-1},A_{i_1+1},
\ldots,A_{i_n-1},A_{i_n+1}  
\dots,A_N} \nonumber \\
&&[\tilde{ \rho}_{A_1+\cdots+A_N}],
\eea
 is a 
$D^n\times D^n$ matrix calculated from \rf{A2}. 
If we denote by $v_i$ 
($i=1,\ldots,D^n$) the eigenvalues of 
$\tilde{\tilde \rho}$ the relations 
\rf{A7}-\rf{A9} give us 
\bea \label{B1}
&&(D^N-1)(\frac{1-x}{D^N})^q + 
(\frac{1+(D^N-1)x}{D^N})^q = \nonumber \\
&&\sum_{i=1}^{D^{n}} 
(\frac{1-x}{D^N}D^{N-n}+xv_i)^q.
\eea
This equation will give us  $x_c(q)$ for each value of $q$. 
Due to the 
monotonic properties of the entropy \rf{A4}, the lowest bound
 for $x_c$ 
is obtained in the limiting case $q \rightarrow \infty$. 
In this limit we can neglect   
 the first term in the left hand side  of \rf{B1} 
and 
keep in the sum only the term corresponding to the largest 
eigenvalue $\bar{v}(N)$ 
of the matrix 
$\tilde{\tilde{\rho}}_{A_{i_1}+\cdots+A_{i_n}}$, which gives
\beq \label{B2}
x_c^S = \frac{1}{1+\frac{D^N}{D^{N-n}-1}(1-\bar{v})},
\eeq
where the superscript refers to entropy. 
In principle we should consider the conditional probabilities 
of arbitrary subsystems. In our applications the largest 
eigenvalue, i.e., the one which produces the most restrictive bound, 
is obtained when we consider, in these conditional 
probabilities,  the subsystem 
$A_1+\ldots +A_{N-1}$.

We are going to consider initially density matrices 
written in vector basis invariant under charge 
conjugation, or spin-reversal symmetry. 
In the $S^z$-basis the basis vectors are 
given by  the non-null combinations 
$|\phi_i^c>= \frac{1}{\sqrt{2}}(|s_1,\ldots,s_N> 
+c|-s_1,\ldots,-s_N>$), ($i=1,\ldots,D^N, c = \pm 1$). 
For $S=\frac{1}{2}$ and $N=2$ this basis recovers 
the standard Bell basis. 
We now consider the density matrix $\rho(x)$ given 
by \rf{A2} where 
$\tilde \rho = |\phi_k^-><\phi_k^-|$ is formed by 
an arbitrary basis vector with charge conjugation 
eigenvalue $c=-1$, i. e., $|\phi_k^->=(|1>-|2>)/
\sqrt{2}$, with $|1> = |s_1,\ldots,s_N>$ and 
$|2> = |-s_1,\ldots,-s_N>$ ($\{s_i\}$ arbitrary). 
It is simple to convince ourselves 
that in the standard basis the only non-zero 
elements of $\tilde \rho$ in \rf{A2} are 
given by
$<1|\tilde \rho|1>=<2|\tilde \rho|2> = 
-<1|\tilde \rho|2>=
-<2|\tilde \rho|1> = 1/2$.
In order to apply the Peres criterion we consider the density 
matrix $\tilde{\sigma}$ defined in \rf{A3} by the transposition of the last 
spin. The only non-zero terms of 
the related matrix 
$\tilde{\sigma}$ are given by
$<1|\tilde {\sigma}|1>=<2|\tilde{\sigma}|2> = 1/2$, 
$<3|\tilde{\sigma}|4>=<1|\tilde{\rho}|2>= 
<3|\tilde{\sigma}|4>=<2|\tilde{\rho}|1>=-1/2$, 
where $|3> = |s_1,\ldots,s_{N-1},-s_N>$ and 
$|4> = |-s_1,\ldots,-s_{N-1},s_N>$. This last matrix has the 
lowest eigenvalue $-1/2$; consequently, from \rf{AA4}, we get the bound
\beq \label{D1} 
x_c = x_c^P = \frac{1}{1+D^N/2},
\eeq
for all values of $D$ and $N$. 

	In the application of the entropic bound, the most 
restrictive condition happens when we consider the conditional 
probability of the subsystem composed by ($N-1$) spins and 
it is not difficult to see that $\tilde{\tilde{\rho}}$ in 
\rf{A9} 
has only two nonzero elements ( $= 1/2$) in the 
diagonal, that gives the eigenvalue $\bar{v} = 1/2$ and from 
\rf{B2} the bound
\beq \label{D2}
 x_c^S = \frac{1}{1 +\frac{D^N}{2}\frac{1}{D-1}}.
\eeq
Comparing \rf{D1} and \rf{D2} we see that for $S=1/2$ both criteria 
give us the same bound $x_c^P =x_c^S= \frac{1}{1+2^{N-1}}$, but for general 
values of the spin $S \neq 1/2$ the bound coming from Peres 
criterion is more restrictive than that of the entropic criterion 
since $x_c^P <x_c^S$. 
It is interesting to remark that in the case $S=1/2$, where the 
bounds coincide the value is known to be the exact value for 
$x_c$, for $N=2$ \cite{horodecki2} or  $N>2$\cite{pittenger}. 
The bounds \rf{D1} 
give us an interesting result. Suppose we need,
 as would occur in a quantum computer, to couple 
the special entangled state $|\phi_k>$ with a  white spectrum 
environment.  The mixed system will be described by the 
density matrix given by \rf{A2}, where the parameter $x$ controls the 
coupling with the environment. The bound \rf{D1} tell us that 
we keep the whole system entangled as long as the ratio between 
the component of our special vector $|\phi_k>$ and any other 
vector in the environment, forming the density matrix, is 
larger than $(1+(D^N-1)x_c^P)/(1-x_c^P)$ that as 
$N\rightarrow \infty$ tend towards 3, independently of the 
value of S. 

A second class of $N$ spins S density matrices we consider 
is a generalization of the Werner density matrices 
\cite{werner}, given 
by  
\rf{A2} with
\beq \label{E1}
|\tilde{\phi}> = \frac{1}{\cal{N}} \sum_{k=-S}^S 
a_k |k,k,\ldots,k>, \;\;\; {\cal{N}} = \sum_{k=-S}^{S} 
|a_k |^2.
\eeq
The standard spin-$S$ Werner density matrix, corresponds to 
$a_{-S}=\cdots =a_S $ and it was shown \cite{pittenger} 
that for this case the exact value where 
quantum entanglement takes 
place is $x_c = 1/(1 +D^{N-1})$. For this particular case it was 
also shown recently \cite{abe2001} 
that the entropic bound \rf{B2} also 
coincides with the above exact value $x_c^S = x_c$. 
The application of the Peres criterion by transposing the last 
spin give us the eigenvalues $\pm |a_l^*a_k|/{\cal{N}}$ 
($l \neq k$) or $|a_k|^2/\cal{N}$ for the matrix 
$\tilde {\sigma}$ in \rf{A3} and consequently the bound 
\rf{AA4} is 
given by
\beq \label{H1}
x_c^P = (1 + D^N \mbox{Max}\{|a_l^*a_k|\} /\sum_{k=-S}^S 
|a_k|^2)^{-1},
\eeq
where we denote by $\mbox{Max}\{|a_l^*a_k|\}$ the maximum value 
of the product $| a_l^*a_k |$ ($l \neq k$) of the coefficients 
forming the general density matrix  \rf{A2} and \rf{E1}. On the other hand 
 the largest eigenvalue 
$\bar {v} = \mbox{Max}\{|a_k|^2\}/\cal{N}$ for the 
matrix $\tilde {\tilde \rho}$ in \rf{C6},  give us, from 
\rf{B2} the entropic bound
\bea \label {H2}
x_c^S &=& (1 +\frac{D^N\mbox{Max}\{|a_l^*a_k|\}}
{\sum_{k=-S}^S |a_k|^2} \alpha)^{-1}, \nonumber \\
\alpha &=& \frac{\sum_{k=-S}^S |a_k|^2 - 
\mbox{Max}\{|a_k|^2\}} 
{(D-1)\mbox{Max}\{|a_l^*a_k|\}}.
\eea
For arbitrary values of $\{a_k\}$ we have 
$\alpha \leq 1$ and from \rf{H1} 
and \rf{H2} we see that the Peres criterion is in general more 
restrictive than the entropic one, i. e., $x_c^P <x_c^S$. 
For the standard Werner density matrix 
($a_{-S} = \cdots = a_S=1$) we have $\alpha =1$ and 
the result $x_c^P= x_c^S = 1/(1+D^{N-1})$, that in this case coincides with 
the exact value of $x_c$ \cite{pittenger}. 
Like in our first 
application \rf{D1} and \rf{D2}, whenever $x_c^P =x_c^S$, 
these values turn out to be the exact one. Equation \rf{H1} and 
\rf{H2} tell us that, whenever at least one of the components $a_k =0$ or 
 at least a pair of coefficients exists such that
$a_k \neq a_{k'}$ in \rf{E1}, $\alpha >1$ and  these bounds are distinct. 
The coincidence of bounds $x_c^P =x_c^S$ happens only for the 
special case $a_{-S} = \ldots =a_S$.

It is important to notice that interpreting \rf{A2} with \rf{E1} as the 
coupling of an entangled state $|\tilde{\phi}>$ with the white spectrum environment, 
in order that the system stays entangled, any component of the environment 
should not exceed a fraction $r_c = (1-x_c)/(1 +(D^N-1)x_c)$ of 
the component of the engineered entangled state $\tilde {\phi}$. 
Using the more restrictive bound $x_c^P$ \rf{H1} we obtain 
$r_c = (\sum_k |a_k |^2 - \mbox{Max} \{|a_l^* a_k |\})/
\mbox{Max} \{ |a_l^* a_k | \} $, that implies that $r_c$ 
decreases with the number of components where $a_k \neq 0$ in \rf{E1} 
and it is more difficult to keep the system 
entangled. 

We have also studied other density matrices where the special entangled 
state $|\tilde{\phi}>$ are invariant under spatial translation  on the spin 
ordering as well charge conjugation. In this case the results have to be carried 
numerically \cite{alctsal} and our results indicate concidences of the 
bounds $x_c^P$ and $x_c^S$ as the number of subsystems $N 
\rightarrow \infty$. 

In conclusion we have shown that, although the Peres transposition method gives us more 
restrictive bounds than those coming from the nonextensive entropy, the 
application of both methods, that in general, at least numerically, is 
not a difficult problem, have the  advantage of possibly providing, when the bounds $x_c^P$ and $x_c^S$ coincide, exact results, that are always of difficult derivation. 
  Also we should remark that our analysis provides 
interesting results that should be useful for quantum computing, teleportation 
or cryptography. If the pure engineered entangled state $|\tilde{\phi}>$ is the 
desired information we are processing ($x=1$ in \rf{A2}),  while time runs, the 
coupling with the environment (represented by $x$ in \rf{A2})  grows ($x$ decreases) 
and the decoerence effect increases. As long as the system stays entangled,  it 
might be possible to recover the desired information with appropriate 
correcting, distillation-like,  procedures. Then, once we know the time dependence of $x(t)$, 
the bounds $x_c^P$ and $x_c^S$ will give us the time scale where the information should 
be processed. 

One of us (C.T.) acknowledges enlightening remarks from R. Horodecki, P. Horodecki and M. Horodecki, as well as warm hospitality at the Schroedinger Institute, where this work was partially performed. 
The present effort has been partially supported by PRONEX, CNPq, and FAPERJ (Brazilian agencies).

\end{multicols}

\begin{thebibliography}{99}

\bibitem{EPR}A. Einstein, B. Podolsky and N. Rosen, Phys. Rev. {\bf 47}, 777 (1935).

\bibitem{schroedinger}E. Schroedinger, Proc. Cambridge Philos. Soc. {\bf 31}, 555 (1935).

\bibitem{GHZ}D.M. Greenberger, M. A. Horne and A. Zeilinger, in {\it Bell's theorem, Quantum Theory , and Conceptions of the Universe}, ed. M. Kafatos (Kluwer Academic, Dordrecht, 1989), p. 73;  D.M. Greenberger, M. A. Horne, A. Shimony and A. Zeilinger, Am. J. Phys. {\bf 58}, 1131 (1990).

\bibitem{ekert}A. Ekert, Phys. Rev. Lett. {\bf 67}, 661 (1991).

\bibitem{bennett}C. Bennett, G. Brassard, C. Crepeau, R. Jozsa, A. Peres and W.K. Wootters, Phys. Rev. Lett. {\bf 70}, 1895 (1993).

\bibitem{zurek2}W.H. Zurek, S. Habib and J.P. Paz, Phys. Rev. Lett. {\bf 70}, 1187 (1993).

\bibitem{quantumchaos}A. Peres, {\it Quantum theory: Concepts and methods}, (Kluwer, Dordrecht, 1993).

\bibitem{popescu}S. Popescu, Phys. Rev. Lett. {\bf 72}, 797 (1994).

\bibitem{barenco}A. Barenco, D. Deutsch, A. Ekert and R. Jozsa, Phys. Rev. Lett. {\bf 74}, 4083 (1995).

\bibitem{zurek}R. Laflamme, C. Miquel, J.P. Paz and W.H. Zurek, Phys.  Rev. Lett. {\bf 77}, 198 (1996).

\bibitem{horodecki1}M. Horodecki, P. Horodecki and R. Horodecki, Phys. Lett. A {\bf 210}, 377 (1996).

\bibitem{horodecki2}M. Horodecki, P. Horodecki and R. Horodecki, Phys. Lett. A {\bf 223}, 1 (1996).

\bibitem{peres}A. Peres, Phys. Rev. Lett. {\bf 77}, 1413 (1996).

\bibitem{horod}P. Horodecki, Phys. Lett. A {\bf 232}, 333 (1997).

\bibitem{horodeckicube}M. Horodecki, P. Horodecki and R. Horodecki, Phys. Rev. Lett. {\bf 78}, 574 (1997). 

\bibitem{popescu2}S. Popescu and D. Rohrlich, Phys. Rev. A {\bf 56}, R3319 (1997).

\bibitem{horodecki3}M. Horodecki, P. Horodecki and R. Horodecki, Phys. Rev. Lett. {\bf 80}, 5239 (1998).

\bibitem{lloyd}H. Touchette and S. Lloyd, Phys. Rev. Lett. {\bf 84}, 1156 (2000). 

\bibitem{bruss}D. Bruss and A. Peres, Phys. Rev. A {\bf 61}, 030301 (2000).

\bibitem{zeilinger}C. Brukner and A. Zeilinger, Phys. Rev. A {\bf 63}, 022113 (2001). 

\bibitem{aberajagopal}S. Abe and A.K. Rajagopal, Physica A {\bf 289}, 157 (2001).

\bibitem{sethmichel}
C. Tsallis, S. Lloyd and M. Baranger, Phys. Rev. A {\bf 63}, 
 042104 (2001); C. Tsallis, P.W. Lamberti and 
D. Prato, Physica A {\bf 295}, 158 (2001).

\bibitem{michalpawel}M. Horodecki and P. Horodecki, Phys. Rev. A {\bf 59}, 4206 (1999).

\bibitem{abe2001}S. Abe, {\it Nonadditive measure and quantum entanglement in a class of mixed states of $N^n$-system}, quant-ph/0104133.

\bibitem{rajagopalrendell}A.K. Rajagopal and R.W. Rendell, {\it Robust and fragile entanglement of three qbits: Relation to permutation symmetry}, quant-ph/0104122.

\bibitem{terhal}B.M. Terhal, {\it Detecting quantum entanglement}, quant-ph/0101032.

\bibitem{tsallis}C. Tsallis, J. Stat. Phys. {\bf 52}, 479 (1988); {\it Nonextensive Statistical Mechanics and its Applications}, 
eds. S. Abe and Y. Okamoto, Series {\it Lecture Notes in Physics} 
(Springer-Verlag, Berlin, 2001). For a regularly updated bibliography see 
http://tsallis.cat.cbpf.br/biblio.htm.

\bibitem{pittenger} A. D. Pittenger and M. H. Rubin, 
Phys. Rev. A {\bf 62}, 032312 (2000).

\bibitem{werner} R. F. Werner, Phys. Rev. A {\bf 40}, 
4277 (1989).

\bibitem{alctsal} F. C. Alcaraz and C. Tsallis, to be published.

\end{thebibliography}
\end{document}